\documentclass{emulateapj}

\shorttitle{{\it Fermi}-LAT Blazar Distributions}
\shortauthors{Singal et al.}

\slugcomment{To Appear in ApJ}

\begin{document}

\title{FLUX AND PHOTON SPECTRAL INDEX DISTRIBUTIONS OF FERMI-LAT BLAZARS AND CONTRIBUTION TO THE EXTRAGALACTIC GAMMA-RAY BACKGROUND}

\author{J. Singal\altaffilmark{1}, V. Petrosian\altaffilmark{1}$^,$\altaffilmark{2}, M. Ajello\altaffilmark{1}$^,$\altaffilmark{3}}

\altaffiltext{1}{Kavli Institute for Particle Astrophysics and Cosmology\\SLAC National Accelerator Laboratory and Stanford University\\382 Via Pueblo Mall, Stanford, CA 94305-4060}
\altaffiltext{2}{Also Departments of Physics and Applied Physics}
\altaffiltext{3}{National Research Council Associate}

\email{jsingal@stanford.edu}

\begin{abstract}
We present a determination of the distributions of photon spectral index and gamma-ray flux - the so called Log$N$-Log$S$ relation -  for the 352 blazars detected with a greater than approximately seven sigma detection threshold and located above $\pm 20^{\circ}$ Galactic latitude by the Large Area Telescope of the {\it Fermi} Gamma-ray Space Telescope in its first year catalog.  Because the flux detection threshold depends on the photon index, the observed raw distributions do not provide the true Log$N$-Log$S$ counts or the true distribution of the photon index.  We use the non-parametric methods developed by Efron and Petrosian to reconstruct the intrinsic distributions from the observed ones which  account for the data truncations introduced by observational bias and  includes the effects of the possible correlation between the two variables.  We  demonstrate the robustness of our procedures using a simulated data set of blazars and then apply these to the real data and find that for the population as a whole the intrinsic flux distribution can be represented by a broken power law with high and low indexes of -2.37$\pm$0.13 and -1.70$\pm$0.26, respectively, and the intrinsic photon index distribution can be represented by a Gaussian with mean of 2.41$\pm$0.13 and width of 0.25$\pm$0.03.  We also find the intrinsic distributions for the sub-populations of BL Lac and FSRQs type blazars separately.  We then calculate the contribution of {\it Fermi} blazars to the diffuse extragalactic gamma-ray background radiation.  Under the assumption that the flux distribution of blazars continues to arbitrarily low fluxes, we calculate the best fit contribution of all blazars to the total extragalactic gamma-ray output to be 60\%, with a large uncertainty.

\end{abstract}


\section{Introduction} \label{intro}

A vast majority of the extragalactic objects observed by the Large Area Telescope (LAT) on the {\it Fermi} Gamma-ray Space Telescope  can be classified as blazars \citep[e.g.][]{Fermiyr1}, a unique subclass of active galactic nuclei (AGNs) for which the jet is aligned with the observer's line of sight \citep[e.g.][]{BK79}.  Analyses of the gamma-ray spectra of blazars along with other signatures of AGNs indicate that the gamma-ray emission is an essential observational tool for understanding of the physics of the central engines of AGNs.  In addition, as in all AGNs, the distribution of spectral and other characteristics of blazars, and the correlations among these characteristics and their cosmological evolutions, are essential information for the studies of  the formation and growth of central black holes of galaxies \citep[e.g.][]{Dermer07}. 

This information comes from the investigation of the population as a whole.  The process for any extragalactic source starts with the determination of the  Log$N-$Log$S$ relation which can be carried out simply by counting sources even before any redshifts are measured and distances are used to determine intrinsic characteristics such as  luminosities and source densities (and their evolution).  Although redshifts are measured for many blazars the extant sample is not yet sufficiently large to allow an accurate determination of the intrinsic characteristics. Our ultimate goal is to carry out such an analysis but the focus of this paper is the determination of the flux and photon spectral index distributions.  

The detection threshold flux of blazars by the {\it Fermi}-LAT depends strongly on an object's gamma-ray spectrum, such that harder spectra are detected at lower fluxes (measured for a given photon energy, here for photons $>100$ MeV). This means that for determination of the flux distribution we need both a measure of flux and the photon index $\Gamma$, and that one deals with a bi-variate distribution of fluxes and indexes, which is truncated because of the above mentioned observational bias.  Thus a bias free determination of the distributions is more complicated than just counting sources.

There have been analyses of this data \citep[e.g.][]{Marco} using Monte Carlo simulations to account for the detection biases. In this paper we use non-parametric methods to determine the distributions directly from the data at hand.  As stressed by Petrosian (1992), when dealing with a bi-(or more generally multi)-variate distribution, the first required step is the determination of the correlation (or statistical dependence) between the variables, which cannot be done by simple procedures when the data is truncated. We use the techniques developed by Efron and Petrosian \citep[EP,][]{EP92,EP99} which can account reliably for the complex observational selection biases to determine first the intrinsic correlations (if any) between the variables and then the mono-variate distribution of each variable. These techniques have  been proven useful for application to many sources with varied characteristics and most recently to radio and optical luminosity in quasars in \citet{Singal11}, where a more thorough discussion and references to earlier works are presented.

In this paper we apply these methods to determine the correlation and the intrinsic distributions of flux and photon index of {\it Fermi}-LAT blazars.   In \S \ref{datasec} we discuss the data used from the LAT extragalactic catalog, and in \S \ref{dandr} we explain the techniques used and present the results.  In \S \ref{bgndradsec}  we describe how the result from such studies are important for understanding the origin of the extragalactic gamma-ray background (EGB) radiation. A brief discussion and summary is presented in \S 5. A test of the procedures using simulated data set is discussed in the Appendix.

\begin{figure}
\includegraphics[width=3.5in]{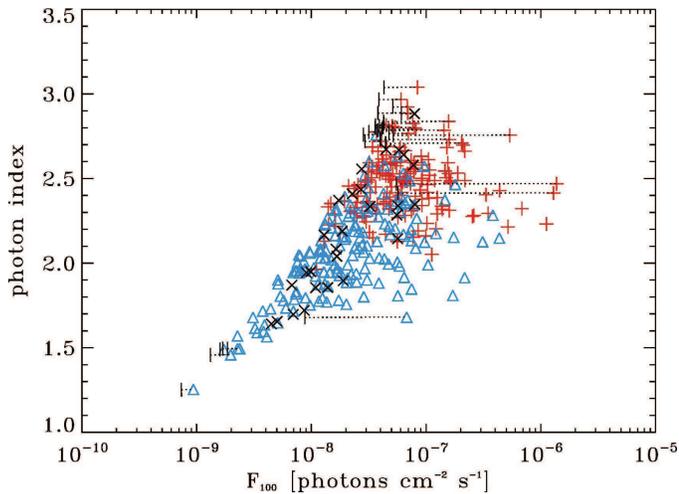}
\caption{ Flux and photon spectral index for the 352 {\it Fermi}-LAT blazars used in this analysis, those with test statistic $\geq$ 50 and $\vert b \vert \geq 20^\circ$.  BL Lac type blazars (n=163) are shown as blue triangles, FSRQs type blazars (n=161) are shown as red plus signs, and blazars of unidentified or ambiguous type (n=28) are represented by black x's.  It is seen that there is a selection bias against soft spectrum sources at fluxes below $\sim 10^{-7}$ photons cm$^{-2}$ sec$^{-1}$.  We also show for a selection of sources (but only a few for clarity) the approximate limiting flux for that source - that is the lowest flux it could have and still be sufficiently bright to be included in the sample given its location on the sky given the reported detection significance.  The location of the line used for the truncation boundary (see \S \ref{corrs}) is shown in Figure \ref{simlumsandsis}.   }
\label{lumsandsis}
\end{figure}

\section{Data}\label{datasec}

In this analysis we use the sources reported in the {\it Fermi}-LAT first year extragalactic source catalog \citep[e.g][]{FermiAGN}.  In particular we rely on the subset of sources that have a detection test statistic $TS \geq 50$ and which lie  at Galactic latitude $\vert b \vert \geq 20^{\circ}$.  This is the same criterion adopted by the LAT team for analysis of the blazar population (\citet{Marco} - hereafter MA) and includes those sources that are fully calibrated and removes spurious sources.  The test statistic is defined as $TS = -2 \, \times \, (\ln (L_0) - \ln (L_1))$, where $L_0$ and $L_1$ are the likelihoods of the background (null hypothesis) and the hypothesis being tested (e.g., source plus background).  The significance of a detection is approximately $n \times \sigma = \sqrt{TS}$ .  Of 425 total such sources, 352 are identified as blazars.  The rest are either identified as radio galaxies (2), other AGN or starbursts (6), high latitude pulsars (9), and objects without radio associations (56).  Among the blazars 161 are identified as Flat Spectrum Radio Quasar (FSRQs) type, 163 are identified as BL Lacertae (BL Lac) type, and 28 have uncertain type.  The fluxes and photon indexes of the blazars are plotted in Figure \ref{lumsandsis}.    

The 352 blazars used in this analysis range in  gamma-ray flux (integrated over the photon energy range 100 MeV to 100 GeV from a power law fit to the {\it Fermi}-LAT data and designated here as $F_{100}$)  from $9.36 \times 10^{-9}$ to $1.37 \times 10^{-6}$ photons cm$^{-2}$ sec$^{-1}$.  The photon index $\Gamma$, obtained by fitting a simple power-law to the spectra in the above energy interval, ranges from  1.253 to 3.039.  The photon index $\Gamma$ is defined such that for the monochromatic photon spectral density $n\!(E)dE \propto E^{-\Gamma}$ (or the $\nu F_{\nu} \propto \nu^{-\Gamma +2}$).  The bias mentioned above is clearly evident; there is a strong selection against soft spectrum sources at fluxes below $F_{100} \sim 10^{-7}$ photons cm$^{-2}$ sec$^{-1}$, caused by the dependence of the {\it Fermi}-LAT point spread function (PSF) with energy \citep{Atwood09}.  

Each source has a $TS$ associated as discussed above, and the background flux is a function of position on the sky, as discussed in \citet{FermiAGN}.  In Figure \ref{lumsandsis} we also show the approximate limiting flux of some  (not all to avoid confusion) objects, an estimate of the lowest flux it should have (at its location in the sky and having the specific value of its index) to be included in the sample, given by $F_{\rm lim}=F_{100} / \sqrt{TS/50}$.  However, as discussed below, because the limiting flux as determined in this way is not the optimal estimate, we use a more conservative truncation as shown by the straight line in Figure \ref{simlumsandsis}. 

\begin{figure}
\includegraphics[width=3.5in]{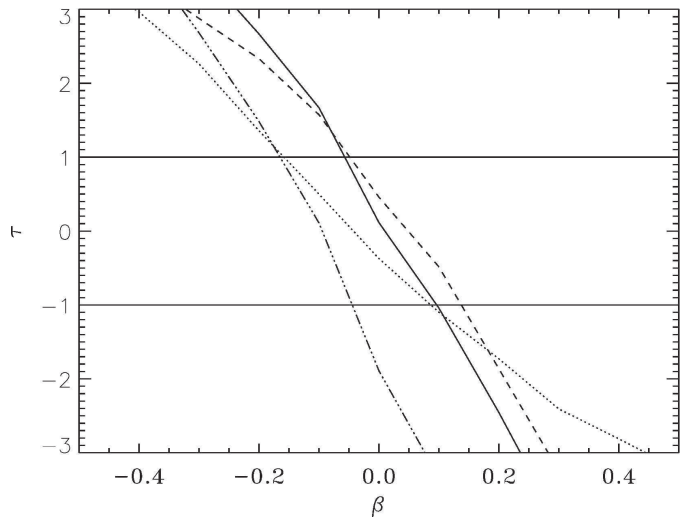}
\caption{ Correlation factor $\beta$ versus test statistic $\tau$ for a photon index and flux correlation of the form given in Equation \ref{fdef}, for the 352 blazars used in this analysis (solid curve), the subset of BL Lac type blazars (dashed curve), and the subset of FSRQs type blazars (dash-dot curve).  The 1$\sigma$ range of best fit values for $\beta$ are where $\vert \tau \vert \leq 1$.  For comparison, the dotted curve shows the correlation factor for just those sources above $5\, \times \,10^{-8}$  photons cm$^{-2}$ sec$^{-1}$, where the data truncation in the $F_{100}$,$\Gamma$ plane is not as relevant. }
\label{corrfig}
\end{figure}

\section{Determinations of Distributions}\label{dandr}

\subsection{Correlations}\label{corrs}

As stressed above, when dealing with a bi-variate truncated data it is imperative to determine whether the variables are independent or not.  If flux and photon index are independent, the combined distribution $G\!(F_{100},\Gamma)$ can be separated into two independent distributions $\psi\!(F_{100})$ and $h\!(\Gamma)$. However, independence may not be the case for $F_{100}$ and $\Gamma$ even though flux is a distance dependent measure while photon index is not.  An intrinsic correlation between the photon index and luminosity may be strong enough to manifest as a flux-index correlation even after cosmological smearing of the correlations, as is the case for example in gamma-ray bursts \citep{Yonetoku04,Lloyd00}.  Even if there is no intrinsic correlation between the photon index and flux, if the selection process introduces some correlation then the independence assumption breaks down, and any such correlation should be removed in order to obtain bias free distributions of the variables.\footnote{The observed distribution (in Figure \ref{lumsandsis}) clearly shows a strong correlation. However, most of this correlation is due to the data truncation described above.}  Thus, the first task is to establish whether the variables are independent. Determining the correlation when the data is truncated is not straight forward. 

We use the Efron-Petrosian method to determine whether the two variables are correlated. This method is a version of the Kendall Tau statistic test devised for truncated data and uses the test statistic

\begin{equation}
\tau = {{\sum_{j}{(\mathcal{R}_j-\mathcal{E}_j)}} \over {\sqrt{\sum_j{\mathcal{V}_j}}}}
\label{tauen}
\end{equation}
to test the independence of two variables in a data set, say ($x_j,y_j$) for  $j=1, \dots, n$.  For untruncated data (i.e. data truncated parallel to the axes) $\mathcal{R}_j$ is the $y$ rank of the data point $j$ within the set with $x_i<x_j$ (or alternatively $x_i>x_j$), which we call the {\it associated set}.  If the data is truncated, say it includes only points with $y>y_{\rm lim}=g(x)$ then the associated set is defined as the largest untruncated set of points associated with $x_j$, i.e. not all points $x_i>x_j$ but only a subset of these that have $y_k \geq y_{{\rm lim},j}=g\!(x_j)$ (see EP for a full discussion of this method). 

If ($x_j,y_j$) were independent then the rank $\mathcal{R}_j$ should be distributed uniformly between 0 and 1 with the expectation value and variance $\mathcal{E}_j=(1/2)(j+1)$ and $\mathcal{V}_j=(1/12)(j^{2}+1)$, respectively.  Independence is rejected at the $n \, \sigma$ level if $\vert \, \tau \, \vert > n$, if $\tau$ turns out to be significantly different than expected value of zero.  In such a cast the correlation is removed parameterically as follows. We define a new variable  $y'=f(x, y)$  and repeat the rank test  with different values of parameters of the function $f$ and determine the nature of the correlation by the best fit value of the parameters  that give $\tau = 0$;  the $n\sigma$ range is obtained from $-n<\tau<n$.

 We carry out this test for our data set using a variable transformation, which is a simple coordinate rotation, by defining a new variable we call the ``correlation reduced photon index'' as

\begin{equation}
\Gamma_{\rm cr}=\Gamma - \beta\, \times\, \log\left({ {F_{100}} \over {F_0}
}\right).  
\label{fdef}
\end{equation}
and determine the value of the parameter $\beta$ empirically that makes $F_{100}$ and $\Gamma_{\rm cr}$ independent, which then means that the distributions of $F_{100}$ and $\Gamma_{\rm cr}$ are indeed separable:

\begin{equation}
G\!(F_{100},\Gamma) = \psi\!(F_{100}) \, \times \, \hat h\!(\Gamma_{\rm cr}). 
\end{equation}
Once the monovariate distributions are determined then the true distribution of $\Gamma$ can be recovered by an integration over $F_{100}$ as:

\begin{eqnarray}
h\!(\Gamma) =  
\nonumber \\
\int_{F_{100}}  \psi\!(F_{100}) \, \hat h\left(\Gamma-\beta \times \log\left({{F_{100}} \over F_0 } \right) \right) \,  d\,F_{100}.
\label{inteq}
\end{eqnarray}
Here $F_0$ is some fiducial flux  we chose to be $F_0=6\times 10^{-8}$ photons s$^{-1}$ cm$^{-2}$ sr$^{-1}$ which is approximately where the flux distribution breaks (see below), although its value is not important.  The data described in \S \ref{datasec} are truncated in the $F_{100}-\Gamma$ plane, due to the bias against low flux, soft spectrum sources.   We can use a curve approximating the truncation, $\Gamma_{\rm lim}=g\!(F_{100})$, which allows us to define the associated sets.  The associated set for each point are those objects whose photon index is less than the limiting photon index of the object in question with its specific value of $F_{100}$. 

We have tested this procedure using a simulated dataset from the {\it Fermi}-LAT collaboration designed to resemble the observations, but with known distributions of uncorrelated photon index and flux and subjected to a truncation similar to the actual data. The results are described in the Appendix where we demonstrate that we can recover the input distributions which are of course quite different than the observed biased distributions. As shown in the Appendix we find that we recover the input distribution best if we start with a truncation boundary $\Gamma_{\rm lim}=g\!(\log F_{100})$ roughly defined by the limiting values obtained from the $TS$ values. We then gradually move this limit to higher fluxes (see Figure \ref{simlumsandsis}) and to more conservative estimations of the truncation. This procedure is stopped when the results do not change significantly. This way we lose some data points but make certain that we are dealing with a complete sample with a well defined  truncation.  Note that when defining new variables the truncation curve as a function of flux should also be transformed  by same parameter $\beta$; 
\begin{equation}
\Gamma_{\rm cr, lim}=\Gamma_{\rm lim} - \beta\, \times\, Log\left({ {F_{100}} \over {F_{100 - min}}}\right).    
\end{equation}

We subject the actual data to the same procedure, and the results converge with the same cutoff limit location.  Figure \ref{corrfig} shows the result of the test statistic $\tau$ as a function of the correlation parameter $\beta$ for a all  blazars and the subsets including only BL Lacs and FSRQs in the sample. Table 1 shows the best fit values and 1$\sigma$ ranges of the correlation parameter $\beta$.  We note that the correlation is weak, for example for all {\it Fermi} blazars the best fit value is $\beta$=0.02 $\pm$ 0.08 indicating a weak correlation with the 1$\sigma$ range including $\beta=0$ or no correlation.

\begin{figure}
\includegraphics[width=3.5in]{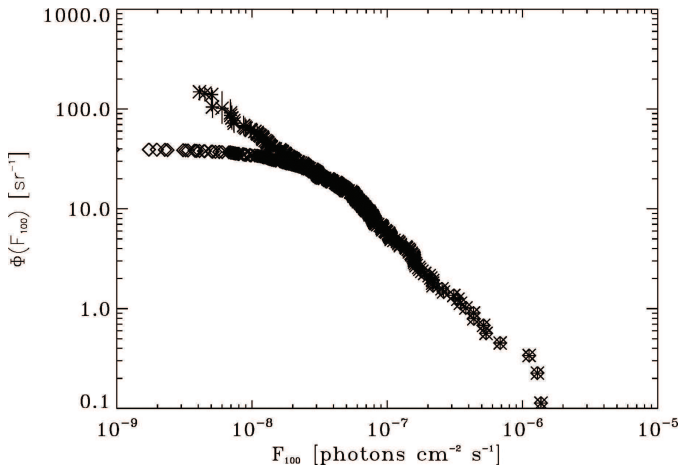}
\caption{ Observed (diamonds) and intrinsic (stars) {\it cumulative} distribution of flux $\Phi\!(F_{100})=\int_{F'_{100}}^\infty \psi\!(F'_{100}) \, dF'_{100}$ for the 352 {\it Fermi}-LAT blazars used in this analysis, shown for the best fit value of the correlation parameter $\beta$. The error bars represent the 1$\sigma$ range of the correlation parameter $\beta$, and are in general smaller than the stars, and are larger than the statistical error.  The normalization is obtained by equation \ref{normeqn}. }
\label{fluxphifig}
\end{figure}

\begin{figure}
\includegraphics[width=3.5in]{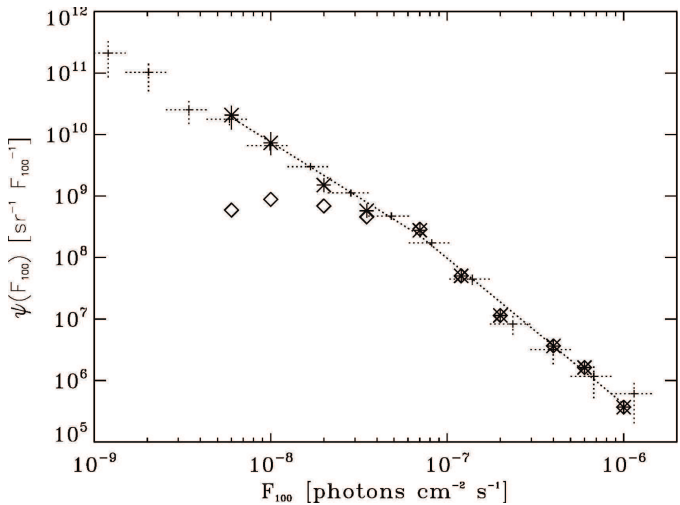}
\caption{ Observed (diamonds) and reconstructed intrinsic (stars) {\it differential} distribution of flux $\psi\!(F_{100})$ for the 352 {\it Fermi}-LAT blazars used in this analysis.  The error bars represent the 1$\sigma$ range of the correlation parameter $\beta$.  The intrinsic distribution is a power law with a break at $F_{\rm br}\simeq 6\, \times \,10^{-8}$  photons cm$^{-2}$ sec$^{-1}$.  The best fit slopes for the intrinsic distribution are -2.37$\pm$0.13 above the break and -1.70$\pm$0.26 below, and the best fit intrinsic distribution is plotted as the dotted line.  We also plot $\psi\!(F_{100})$ as determined in MA (small crosses), with error bars (dotted lines). The best fit value for $\psi\!(F_{break})$ is 2.2 $\times 10^8$ sr$^{-1} F_{100}^{-1}$. }
\label{fluxdistfig}
\end{figure}

\begin{figure}
\includegraphics[width=3.5in]{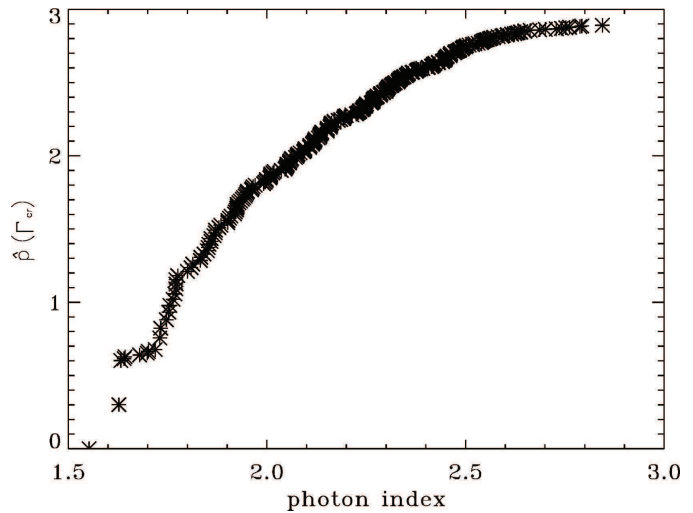}
\caption{Intrinsic cumulative distribution of photon index $\hat P\!(\Gamma_{\rm cr})=\int_0^{\Gamma_{\rm cr}} \hat h\!(\Gamma_{\rm cr}) \, d\Gamma_{\rm cr}$ for the 352 {\it Fermi}-LAT blazars used in this analysis. The normalization of $\hat P$ is arbitrary }
\label{aphifig}
\end{figure}

\begin{figure}
\includegraphics[width=3.5in]{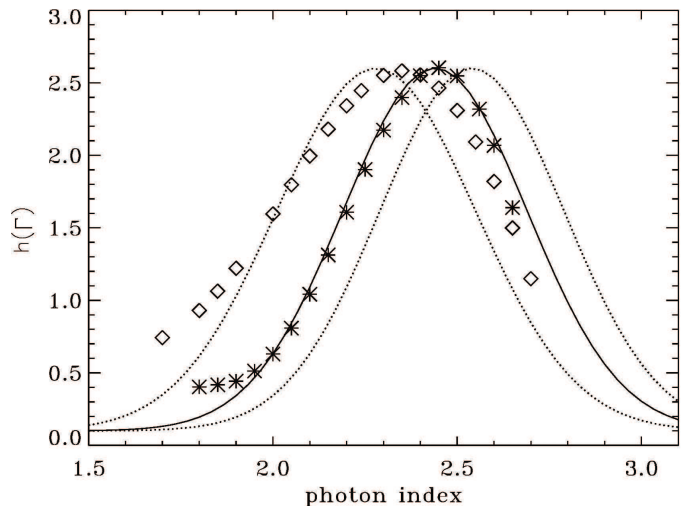}
\caption{Observed (diamonds) and reconstructed intrinsic (stars) distribution of photon index $h\!(\Gamma)$ for the 352 {\it Fermi}-LAT blazars used in this analysis. The intrinsic distribution is calculated from the flux distribution and the correlation reduced photon index distribution by equation \ref{inteq}.  The stars represent the intrinsic distribution calculated with the best fit value of the correlation parameter $\beta$ and the solid curve is the best fit Gaussian function to these values, while the dotted curves represent the best fit Gaussian functions to the extremal intrinsic distributions allowed by the 1$\sigma$ range of $\beta$.  The intrinsic distribution can be represented by a Gaussian with a mean of 2.41$\pm$0.13 and 1$\sigma$ width of 0.25$\pm$0.03, while the observed distribution can be represented by a Gaussian with a mean of 2.32$\pm$0.01 and 1$\sigma$ width of 0.32$\pm$0.01.  The normalization of $h\!(\Gamma)$ is arbitrary.}
\label{adistfig}
\end{figure}

\newpage

\subsection{Distributions}\label{dist}

With the correlation removed the independent distributions $\psi\!(F_{100})$ and $\hat h\!(\Gamma_{\rm cr})$ can be determined using a method outlined in \citet{P92} and developed by \citet{L-B71}.  These methods give the cumulative distributions by summing the contribution from each point without binning the data. 

\subsubsection{flux distributions}\label{fdists}

For the flux the cumulative distribution

\begin{equation}
\Phi\!(F_{100}) \equiv \int_{F_{100}}^\infty {\psi\!(F_{100}') \, dF_{100}'}
\end{equation}
is obtained as

\begin{equation}
\Phi\!(F_{100}) = \prod_{j} \left({1 + {1 \over N\!(j)}} \right)
\label{phieq}
\end{equation}
where $j$ runs over all objects with fluxes $F_{100,j}\geq F_{100}$, and $N(j)$ is the number of objects in the {\it associated set} of object $j$; namely those with  with a value of $F_{100,i} \geq F_{100,j }$ and  $\Gamma_{\rm cr,i} \leq \Gamma_{\rm cr, lim}(F_{100,j})$ determined from the truncation curve described above.  Equation \ref{phieq} represents the established Lynden-Bell method.  We note again that the cutoff curve as a function of flux is scaled by $\beta$ in the same manner of equation \ref{fdef}. The use of only the associated set for each object removes the biases introduced by the truncation.  

The differential distribution 
\begin{equation}
\psi\!(F_{100}) = - {d \Phi\!(F_{100}) \over dF_{100}}
\label{psieqn}
\end{equation}
 is obtained by fitting piecewise polynomial functions via least-squares fitting to $\Phi\!(F_{100})$ and calculating its derivative.  Figure \ref{fluxphifig} shows the true intrinsic and (the raw) observed cumulative distributions of $F_{100}$ for all 352 blazars, while figure \ref{fluxdistfig} shows the calculated true intrinsic differential distribution $\psi\!(F_{100})$, along with those obtained from the raw observed data without correcting for the bias.  A direct comparison to the results from MA is presented there as well.   The differential counts manifest a broken power law which can be fit by the form

\begin{eqnarray}\label{bpowlaw}
\psi\!(F_{100}) = \,\,\,\,\, \psi\!(F_{break}) \, \left( {  {{F_{100}} \over {F_{break}}} } \right)^{m_{above}}  \,\,\,\,\,{\rm for} \,\,\,\,\,F_{100} \geq F_{break} \\
\nonumber  \psi\!(F_{break}) \, \left( {  {{F_{100}} \over {F_{break}}} } \right)^{m_{below}} \,\,\,\,\,{\rm for} \,\,\,\,\,F_{100}< F_{break}.
\end{eqnarray}
$m_{above}$ and $m_{below}$ are the power law slopes above and below the break, respectively, and are obtained from a least-squares fitting of $\psi\!(F_{100})$, as is the value of $F_{break}$.  At values of $F_{100}$ above $F_{NT} \equiv 1 \times 10^{-7}$  photons cm$^{-2}$ sec$^{-1}$ the truncation is not significant and we can obtain the normalization by scaling the cumulative distribution $\Phi\!(F_{100})$ such that.  

\begin{equation}
\Phi\!(F_{NT}) = { {N \pm \sqrt{N}} \over {8.26 \, {\rm sr} } } , 
\label{normeqn}
\end{equation}
where $N$ is the number of objects above  $F_{NT}$ and is equal to 60 for all blazars, 12 for BL Lacs, and 48 for FSRQs. The $\sqrt{N}$ uncertainty arises because of Poisson noise for the brightest sources, and 8.26 sr is the total sky coverage considered, which is $\vert b \vert \geq 20^{\circ}$ as discussed in \S \ref{datasec}.  This also gives the value of $\psi\!(F_{break})$ by inegrating $\psi\!(F_{100})$ at fluxes above $F_{NT}$ and setting this equal to $\Phi\!(F_{NT})$:

\begin{equation}
\psi\!(F_{break}) = \Phi\!(F_{NT}) \, (m_{above}-1)  \,  F_{NT}^{-m_{above}-1} \, F_{break}^{m_{above}}.
\label{normeqn2}
\end{equation}
The best fit value for $\psi\!(F_{break})$, corresponding to the best-fit value of $m_{above}$, is then 2.2 $\times 10^8$ sr$^{-1} F_{100}^{-1}$. 

\subsubsection{Photon index distributions}\label{pdist}

A parallel procedure can be used to determine the distribution of the correlation reduced photon index, namely the cumulative distribution 

\begin{equation}
\hat P\!(\Gamma_{\rm cr}) \equiv \int_0^{\Gamma_{\rm cr}} {\hat h\!(\Gamma_{\rm cr}') \, d\,\Gamma_{\rm cr}'}
\end{equation}
obtained with

\begin{equation}
\hat P\!(\Gamma_{\rm cr}) = \prod_{k} \left({1 + {1 \over M\!(k)}} \right)
\end{equation}
in the method of \citet{L-B71} can be differentiated to give the differential distribution

\begin{equation}
\hat h\!(\Gamma_{\rm cr}) = {d \hat P\!(\Gamma_{\rm cr}) \over d\,\Gamma_{\rm cr}}
\end{equation}

In this case, $k$ runs over all objects with a value of $\Gamma_{\rm cr, k}\leq \Gamma_{\rm cr}$, and $M(k)$ is the number of objects in the {\it associated set} of object $k$; i.e. those with $\Gamma_{cr, i} \leq \Gamma_{cr, k}$  and $F_{100}m \geq F_{\rm lim,k}$ obtained from the truncation line at $\Gamma_{\rm cr,k}$.  

Figure \ref{aphifig} shows the cumulative distribution of the correlation reduced photon index ${\hat P}\!(\Gamma_{\rm cr})$ for all 352 blazars.  Differentiation of this gives ${\hat h}\!(\Gamma_{\rm cr})$, which can be substituted in equation \ref{inteq} to obtain the intrinsic distribution of the photon index itself, $h\!(\Gamma)$. The results are shown in Figure \ref{adistfig} along with the raw observed distribution for comparison.  Because the mean of intrinsic distribution of photon index is sensitive to the value of the correlation parameter $\beta$, we include the full range of intrinsic distributions resulting from the 1$\sigma$ range of $\beta$. A Gaussian form provides a good description of the intrinsic distribution of the index.  

We have carried out  identical procedures to obtain the distributions of the BL Lac and FSRQ subsets of the data.  Table 1 summarizes the best fit parameters for the intrinsic flux and photon index distributions, for the sample considered as a whole, and for the BL Lac and FSRQs sub-populations separately.  The errors reported include statistical uncertainties in the fits and the deviations resulting from the 1$\sigma$ range of the correlation parameter $\beta$.  A higher value of $\beta$ (i.e. more positive correlation between flux and photon index absolute value) moves the mean of the photon index distribution down to a lower absolute value of the photon index  and makes the faint end source counts slope less steep (less negative $m_{below}$), while a lower value of $\beta$ has the opposite effect.

\subsection{Error Analysis}

It addition to uncertainty in the value of $\beta$ and those due to the fitting procedure there are other effects that can add to the uncertainties of the final results. Here we consider the effects of some factors which we have ignored in the above analysis.

1. {\it Individual measurement uncertainties}:  We have treated individual sources as points having a delta function distribution in the flux-index plane, resulting in a possible Eddington bias \citep{Eddington40}. The measurement uncertainties can be included by changing the delta functions to kernels whose widths are determined by the reported measurement errors.  The main effect of this will be smearing out of the distribution which can be neglected if the errors are small compared to the width of features in the distributions. This effect will not introduce any bias as long as measurement uncertainties are symmetrical about the reported value (e.g the kernels are Gaussian) and the distributions themselves are symmetrical or fairly flat.  The former is the case for the reported uncertainties in \citet{Fermiyr1}.  This later is the case for the distribution of $\Gamma$, where the reported measurement errors vary between 0.04 and 0.35, but this is not likely to introduce any bias given the symmetrical distribution in $\Gamma$. 

The situation is different for the flux distribution, which is a power law. In this case the typical flux uncertainty values are on the order of 1/4-1/3 of the reported fluxes. There may be more or fewer sources be included than missed in a flux limited sample such as this one.  For example, for a power law differential distribution with index $|m_{below}|>1$, which is the case here, more sources will be included than missed at any flux for which there are errors in the reported fluxes, which will bias the distribution.  The effect is different for the case where fractional measurement errors are constant with flux versus when they change with flux.  For constant fractional flux measurement errors, an error will be introduced on the normalization of the source counts and can be approximated by [$1/2 \, \delta^2 \, m_{below} \, (m_{below}+1)$] \citep{Teer04} where $\delta$ is the fractional error in flux.  For $m_{below} \sim-1.7$ and $\delta \sim 0.3$ this error will be about 5\%, which is small compared with other sources of normalization uncertainty.  On the other hand, there will be an effect on the reconstructed slope of the counts only if the fractional flux measurement errors change with flux.  It is expected that the fractional flux errors will be larger for lower fluxes, and in the extreme case that they do increase from negligible at high fluxes to values of 1/4-1/3 at the lowest fluxes, according to simulation results in \citet{Teer04} resulting fractional errors in the source count slope will be around 7\%, which is significantly smaller than the errors we already quote for the source count slopes.   

\citet{CP93} evaluated the effect of measurements error on EP method determinations of luminosity functions of quasars using a Gaussian kernel and found that for individual uncertainty widths significantly smaller than the data range, the effects of the inclusion of measurement uncertainties were small (e.g. Figure 1 of that work).  Given the Gaussian symmetrical nature of the reported uncertainties, the symmetrical nature of the photon index distribution, and the relatively shallow faint end power law slope of the source counts distribution, and especially the relative size of the reported uncertainties compared to the range of values considered, in light of the analysis done by Caditz and Petrosian we consider the errors introduced by the individual data point uncertainties to be negligible compared to the uncertainties introduced by the range of the correlation parameter $\beta$ and the uncertainties in the power-law fits.  

2. {\it Blazar variability}:  It is well known that blazars are inherently variable objects.  There are two potential effects arising from blazar variability relevant to the analysis here.  One is similar to measurement error discussed above, in that it presumably would cause more objects to rise above the flux limit and be included in the survey than go below the flux limit and be excluded.  The other is that the reported temporally averaged quantities such as flux and photon index, which we use in this analysis, may deviate from the true average values.   

Addressing the former issue, as discussed in the first year {\it Fermi}-LAT extragalactic source catalog \citep[e.g. Figure 11c of][]{FermiAGN}, the pattern of maximum flux versus mean flux does deviate at the lowest detected fluxes.  This indicates that variability becomes more important with decreasing flux, but not as sharply as might be expected from previous EGRET data.  This will be even less sharp for the $TS \geq 50$ sources we use as opposed to the entire $TS \geq 25$ sample considered there.  As shown in \citet{Fermiflux}, the peak-to-mean flux ratio is a factor of two or less for most {\it Fermi}-LAT blazars, which excludes the possibility that most of the sources are detected because of a single outburst which happened during the 11 months of observation and are undetected for the remaining time.  We believe the bias resulting from detecting blazars only in their flaring state is small.

Addressing the later issue, both \citet{Fermivar1} and \citet{Fermivar2} presented a detailed analysis of the variability issues with {\it Fermi-LAT} blazars.  They find that most sources exceed their average flux for less than 20\%, and often less than 5\%, of the monitored time, and conclude that both the timescale of variability is short compared with the length of observations, and that the measured average quantities are not highly biased by flaring.  Moreover, as also shown in \citet{Fermiflux}, there is little or no temporal variation of the photon index with flux.  We thus believe that no large systematic uncertainties result from the use of these averaged physical quantities.\footnote{There is an interesting implication in blazar variability for the extragalactic gamma-ray background, which is discussed in \S \ref{bgndradsec}.}

3. {\it Source confusion}:  Source confusion can also introduce errors at the faint end of the reconstructed distributions because of relatively broad point spread function of the {\it Fermi}-LAT;  some faint sources may be either missed entirely or erroneously combined into the fluxes of other sources.  We first note that these two phenomena will have opposite systematic effects on the faint end source counts slope, as the former would tend to make it less steep while the later would tend to make it steeper.  In addition to this self-canceling tendency, several tests argue against {\it Fermi}-LAT's blazar detections being significantly confusion limited. \citet{Fermiyr1} estimates that at Galactic latitudes above $\pm 10^\circ$ and at a $TS \geq 25$ detection threshold, approximately 7.6\% of sources (80 out of 1043) are missed because of confusion, and blazars are 85\% of the $\vert b \vert \geq 10^{\circ}$ sources.  Since we have used only sources detected at $TS\geq 50$ and at latitudes above $\pm 20^\circ$ Galactic latitude, the effect of confusion should be lower because the sample will be more complete.  Again, the faint end source counts slope could be altered by an amount considerably less than this.  Other evidence against source confusion being a significant problem for {\it Fermi}-LAT blazars is the large increase in the number of extragalactic sources from the first-year to second-year {\it Fermi}-LAT catalogs \citep[e.g][]{Fermivar2}.  Additionally, there is the analysis described in section 8.3 of MA where different extragalactic sky scenarios were simulated and run through an instrument detection and catalog pipeline, including an extremal scenario with a single steep power law distribution of blazars with a differential slope of -2.23, in which blazars with fluxes greater than 10$^{-9}$ photons cm$^2$ s$^{-1}$ would produce 70\% of the total extragalactic gamma-ray background.  Even under this much more dense sky scenario, many more blazars would be detected by the instrument and analysis, including at low fluxes.  

We also note that to the extent that the {\it effects} of both blazar variability and source confusion will have some photon index dependence, because of differing spectra in the case of the former and the {\it Fermi}-LAT’s energy-dependent point spread function in the case of the later, then any potential biases will already have been accounted for in the way we have dealt with truncations in the $F_{100},\Gamma$ plane.  We have treated the truncation in that plane as empirical and accounted for it as discussed in \S \ref{corrs}.  

In summary, most of these additional sources of error are small and are important only for a small range of fluxes around the lowest fluxes, where the EP method has a larger uncertainty anyway, and for that reason these low fluxes are excluded from the fitting in our analysis (compare the end points of raw and corrected distributions in Figures 3 and 4).

\section{Total Output from Blazars and the Extragalctic Gamma-ray Background}
\label{bgndradsec}

\begin{figure}
\includegraphics[width=3.5in]{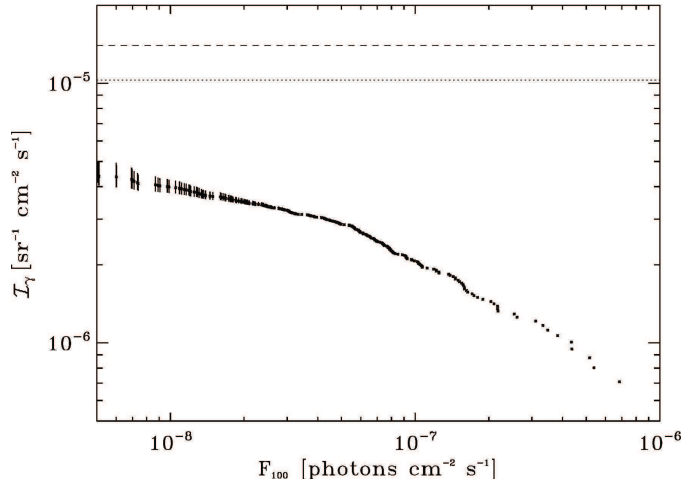}
\caption{ Estimate of the cumulative number of photons between 0.1 and 100 GeV above a given $F_{100}$ from blazars, $\mathcal{I}_{\rm \gamma - blazars}(>F_{100})$ from equation \ref{CGB2}, shown with error bars resulting from the 1$\sigma$ range of the correlation parameter $\beta$.  The bottom dotted horizontal line shows the level of the EGB as measured by Fermi \citep{Fermibgnd}, with {\it Fermi} resolved point sources removed.  The top dashed horizontal line shows the EGB, ie the total extragalactic gamma-ray output ($\mathcal{I}_{EGB}$) as defined here.  The best fit total contribution of blazars to $\mathcal{I}_{EGB}$, obtained by integrating equation \ref{CGB2} to zero flux, is 59\%, but our analysis cannot rule out blazars, integrated to arbitrarily low flux, forming as little as 33\%  or as much as  all of the total  extragalactic gamma-ray output, due to the large range of uncertainty in the faint end source counts slope, ultimately owing to the uncertainty in the correlation parameter $\beta$.  }
\label{bgndcontfig}
\end{figure}

We can use the above results to calculate the total flux from blazars and the contribution of blazars to the EGB, defined here as the total extragalactic gamma-ray photon output.\footnote{This definition avoids the problem that individual instruments resolve a different fraction of sources, leading to different estimates for the fraction of the total extragalactic photon output that is unresolved.}  The total output in gamma-ray photons from blazar sources with fluxes greater than a given $F_{100}$, in terms of photons s$^{-1}$ cm$^{-2}$ sr$^{-1}$ between 0.1 to 100 GeV, is 

\begin{equation}
\mathcal{I}_\gamma(>F_{100})=\int_{F_{100}}^{\infty} \, F'_{100} \, \psi\!(F'_{100}) \, dF'_{100}.
\label{CGB1}
\end{equation}
Integrating by parts the contribution to the EGB can be related directly to the cumulative distribution $\Phi\!(F_{100})$ which is the primary  output of our procedure
\begin{equation}
\mathcal{I}_\gamma(>F_{100})=F_{100} \, \Phi\!(F_{100})+\int_{F_{100}}^{\infty} \Phi\!(F'_{100}) \, dF'_{100}.
\label{CGB2}
\end{equation}
The advantage of using the latter equation is that it can give a step-by-step cumulative total contribution to the background instead of using analytic fits to the differential or cumulative distributions obtained from binning the data.  Figure \ref{bgndcontfig} shows $\mathcal{I}_\gamma(>F_{100})$  resulting from this integration down to flux  $F_{100}= 5\times 10^{-9}$, with the total output of the blazar population {\it at fluxes probed by this analysis}  being $\mathcal{I}_{\rm \gamma}(>F_{100}=5 \times 10^{-9})$=4.5 $\pm$ 0.5 $\times 10^{-6}$ photons s$^{-1}$ cm$^{-2}$ sr$^{-1}$.   
Note that this includes the contribution from both detected blazars and those undetected above this flux owing to the truncation in the $F_{100}, \Gamma$ plane.  Therefore, as expected it is more than the total contribution of blazars resolved by {\it Fermi}-LAT which is estimated to be $4.1\pm 0.2\times 10^{-6}$ photons s$^{-1}$ cm$^{-2}$ sr$^{-1}$ \citep{FermiAGN}.\footnote{Actually the latter is what is attributed to point sources with test statistic value of $TS>25$ which corresponds roughly to a 5$\sigma$ detection.}

In order to determine the contribution of  blazars with $F_{100}< 5\times 10^{-9}$ photons s$^{-1}$ cm$^{-2}$ sec$^{-1}$ to the total EGB we must  extrapolate the flux distribution we have obtained to below this flux which cannot be unique and is more uncertain. We fit a power law to the faint end of $\Phi\!(F_{100})$  so that we can extend the integration of equation \ref{CGB2} to lower fluxes. Extending to zero flux we find that blazars in toto can produce a photon output of $\mathcal{I}_{\gamma}(>F_{100}=0)$=8.5 (+6.3/-2.1) $\times 10^{-6}$ photons s$^{-1}$ cm$^{-2}$ sr$^{-1}$.  This large range is due to the uncertainty in the faint end cumulative source counts slope, ultimately owing to the range of the correlation parameter $\beta$, where the best fit value reported is for the middle of the 1$\sigma$ faint end slope of $\Phi\!(F_{100})$. 

This is to be compared with the total observed EGB of $\mathcal{I}_{EGB}=14.4 \pm 1.9 \times 10^{-6}$ photons s$^{-1}$ cm$^{-2}$ sr$^{-1}$
reported by {\it Fermi}.\footnote{The {\it Fermi} collaboration papers divide this radiation into two parts, one from what is referred to as the contribution of resolved sources (the $\mathcal{I}_{sources}=4.1 \pm 0.2 \times 10^{-6}$ photons s$^{-1}$ cm$^{-2}$ sr$^{-1}$ mentioned above), and a second ``diffuse'' component of $\mathcal{I}_{EGB-sources}=1.03 \pm 0.17 \times 10^{-5}$ photons s$^{-1}$ cm$^{-2}$ sr$^{-1}$ \citet{Fermibgnd}. However, the most relevant comparison is with the total of these two, because which sources are declared to be resolved is determined by a $TS$ threshold, not a flux limit, and these are different due in part to the truncation in the $F_{100},\Gamma$ plane and the varying Galactic diffuse level.}  If the blazar population continues to have the fitted power law distribution to zero flux then it is clear that for our best fit parameters blazars can produce 59\% of the observed EGB but this contribution could be as little as 39\% or as much as all of the total extragalactic gamma-ray output of the Universe.  This result is in agreement, albeit with a larger uncertainty, with the result in MA, where following the definition conventions here blazars extrapolated to zero flux are found to contribute 46\% $\pm$ 10\% of the EGB.\footnote{The total point source diffuse emission and EGB intensity presented in Table 6 of MA have the contribution of resolved {\it Fermi}-LAT sources removed, so for direct comparison to the results presented here the total {\it Fermi}-LAT resolved source contribution of $4.1\pm 0.2\times 10^{-6}$ photons s$^{-1}$ cm$^{-2}$ sr$^{-1}$ must be added to both before the ratio is taken.}  

It is, however, likely that blazars do not continue as a population with no change in the source counts slope to zero flux, since even a dim AGN of luminosity $\sim$10$^{45}$ erg s$^{-1}$ at redshift 3 would have a flux of $\sim$10$^{-12}$ photons cm$^{-2}$ sec$^{-1}$.  If we only integrate equation \ref{CGB2} to this lower flux limit, then we get the total blazar contribution to be $\mathcal{I}_{\gamma}(>F_{100}=1 \times 10^{-12})$=7.7 (+0.8/-1.2) $\times 10^{-6}$ photons s$^{-1}$ cm$^{-2}$ sr$^{-1}$, which brings the upper limit estimate down to 66\% of the total EGB.

We can also obtain the energy intensity of the cumulative emission from blazars as 
\begin{equation}
I_{\rm \gamma - blazars}(>F_{100})=\int_{F_{100}}^\infty dF'_{100}\, \int_{-\infty}^\infty d\Gamma  E\!(F'_{100}, \Gamma) \, G\!(F'_{100}, \Gamma)
\end{equation}
where for a simple power law spectrum we can relate the energy emitted between 0.1 and 100 GeV to the flux as
\begin{equation}
{ {E_{100}} \over {F_{100} } }\equiv R(\Gamma) \cong 100 \, \times \, {{\Gamma-1} \over {\Gamma-2}} \times {{1-10^{3(2-\Gamma)}} \over {1-10^{3(1-\Gamma)}} } \,\,\,{\rm MeV / photon},
\label{econv}
\end{equation}
except for $\Gamma=2$ and $\Gamma=1$ for which $R(2)=\ln 10^3/(1-10^{-3})\sim 6.9$ and  $R(1)=(10^3-1)/\ln 10^3\sim 150$ respectively. If we ignore the weak correlation between $\Gamma$ and $F_{100}$ (set $\beta=0$) we get $I_{\rm \gamma - blazars}(>F_{100})={\bar R}\times \mathcal{I}_{\rm \gamma}(>F_{100})$ where ${\bar R}$ is average value of $R$ over the Gaussian distribution of $\Gamma$. We carry this average numerically and get the total (resolved and unresolved) $I_{\rm \gamma - blazars}=2.7 (+3.1/-0.9) \times 10^{-3}$ MeV cm$^{-2}$  sec$^{-1}$ sr$^{-1}$, integrating to zero flux taking into account the uncertainties above and the 1$\sigma$ uncertainty in the mean of $h\!(\Gamma)$.

We note that this analysis can not rule out blazars as the sole significant contributor to the EGB, although the best fit value does not favor this being the case.  The spectral index of the EGB of $\sim$2.4 \citep{Fermibgnd} is consistent with the mean photon index of the blazars as determined here and in MA.  In a similar vein, \citet{VP11} have shown that the spectrum of the EGB is consistent with a blazar origin.  Several authors \citep[e.g.][]{SV11,A10} have suggested that blazars could be the primary source of the EGB, while the results presented in MA and \citet{MH11} would favor the primary source being something else.  Other possible source populations include starforming galaxies, which have been recognized as a possible major contributor to the EGB by e.g. \citet{SV11}, \citet{Fields10}, and \citet{L11}, although this has been disputed by \citet{M11}, radio galaxies \citep[e.g][]{I11}, and other non-blazar AGN \citep[e.g.][]{IT09,IT11}.

According to \citet{SS96}, to the extent that faint blazars are more likely to be observed by instruments such as the {\it Fermi}-LAT if they are in the flaring state rather than the quiescent state, then the observed blazars should have a different mean mean photon index than the EGB, were the EGB to be made primarily from quiescent state blazars, under the assumption that blazars in the flaring state have a different spectrum than in the quiescent state.  As the reconstructed mean photon index here of the {\it Fermi}-LAT observed population is close to that of the EGB, and there is only a weak relation and correlation between flux and photon index, this would imply that at least one of the following must be the case: a) there is not a significant bias in the {\it Fermi}-LAT toward detecting blazars in the flaring state, b) quiescent blazars do not form the bulk of the EGB, or c) flaring and quiescent blazars have, en masse, roughly the same photon index distributions. 

\section{Discussion}\label{discsec}

We have used a rigorous method to calculate the intrinsic distributions in flux (known commonly as the source counts or the Log$N$-Log$S$ relation) and photon index of {\it Fermi}-LAT blazars directly from the observed ones without any assumptions or reliance on extensive simulations.  This method features a robust accounting for the pronounced data truncation introduced by the selection biases inherent in the observations, and addresses the possible  correlation between the variables. The accuracy of the methods used here are demonstrated in the Appendix using a simulated data set with known distributions. A  summary of the best fit correlations between photon index and flux, and the best fit parameters describing the inherent distributions of flux and photon index, of the observed data are presented in Table 1 along with the values obtained by MA.  We have obtained the distributions of flux and photon index of blazars considering the major data truncation arising from {\it Fermi}-LAT observations.  More subtle issues affecting the distributions we have derived, especially the photon index distribution, may arise due to the finite bandwidth of the {\it Fermi}-LAT and lack of complete knowledge of the objects' spectra over a large energy range and deviations from simple power laws.  However the {\it Fermi}-LAT bandwidth is sufficiently large that the contribution of sources which peak outside of this range to the source counts and the EGB in this energy range will be small. 

We find that the photon index and flux show a slight correlation, although this correlation is of marginal significance. This indicates that the intrinsic luminosities and photon indexes are correlated only weakly.  The comparison of the intrinsic and raw observed distributions show clearly the substantial effects of the observational bias.  The intrinsic differential counts can be fitted adequately by a broken power law and the photon index appears to have a intrinsic Gaussian distribution. We also find that in general the values reported here are consistent with those reported in MA for the power law slopes of the flux distribution $\psi\!(F_{100})$ and the distributions of photon index $h\!(\Gamma)$, although the allowed range of the correlation parameter $\beta$ here allows for wider uncertainty in these values in some cases.  We do note a discrepancy at the 1$\sigma$ level for the faint end slope of the FSRQ source counts.  

Using the bias free distributions we calculated the total cumulative contribution of blazars to the EGB as a function of flux. We obtain this directly from the cumulative flux distribution which is the main output of the methods used.  Under the assumption that the distribution of blazars continues to arbitrarily low flux, we find the best fit contribution of blazars to the total extragalactic gamma-ray radiation in the range from 0.1 to 100 GeV to be at the level of 59\%, although this analysis cannot rule out blazars producing as little as 39\% or as much as all of the total extragalactic gamma-ray output.  This result is in agreement with the result in MA, although with a larger uncertainty.  The significant uncertainties reported here for the source count slopes and the contribution of blazars to the EGB are ultimately due to the allowed range of the correlation parameter $\beta$.  As discussed in the Appendix, the method applied to the (uncorrelated) simulated data also manifests a significant uncertainty on $\beta$, that translates into the corresponding uncertainty for the faint-end slope of the source count distribution. This is important as it ultimately governs the contribution of blazars to the EGB.  We note that a similar scenario  (i.e. absence of a correlation between photon index and flux) might characterize the real data in view of the results reported in the previous sections and their similarity to the results obtained using simulated uncorrelated data.  As shown in the Appendix, larger samples are required to narrow down the uncertainty on the correlation parameter $\beta$.

If, as could be expected, the flux distribution flattens at fluxes below $\sim$10$^{-12}$ photons cm$^{-2}$ sec$^{-1}$, the integrated contribution will be significantly lower than for a naive extrapolation to zero flux.  This is also modulo any change in the power law slope of the source counts below the fluxes probed in this analysis, which might arise due to luminosity and/or density evolution with redshift.  A full accounting for the possible evolution in the blazar population using a sample with redshift determinations will be presented in a forthcoming work.

\acknowledgments

The authors thank the members of the {\it Fermi} collaboration.  JS thanks S. Kahn and R. Schindler for their encouragement and support.  VP acknowledges support from NASA-Fermi Guest Investigator grant NNX10AG43G.

\begin{deluxetable}{lccccccc}\label{tizzable}
\tabletypesize{\scriptsize}
\tablecaption{Best fit parameters for Fermi-LAT blazar intrinsic distributions, as calculated in this work and in MA \citep{Marco}} 
\tablecolumns{8}
\startdata
  & n & $\beta$\tablenotemark{a}  & $m_{above}$\tablenotemark{b} & $F_{break}$\tablenotemark{c} & $m_{below}$\tablenotemark{d} & $\mu$\tablenotemark{e} & $\sigma$\tablenotemark{f} \\
\hline
\\
Blazars\tablenotemark{g} (this work) & 352 & 0.02$\pm$0.08 &-2.37$\pm$0.13 & 7.0$\pm$0.2 & -1.70$\pm$0.26 & 2.41$\pm$0.13 & 0.25$\pm$0.03 \\
Blazars\tablenotemark{g} (MA)        & 352 & -             & -2.48$\pm$0.13 & 7.39$\pm$1.01 & -1.57$\pm$0.09 & 2.37$\pm$0.02 & 0.28$\pm$0.01 \\
\\
BL Lacs (this work)                  & 163 & 0.04$\pm$0.09 & -2.55$\pm$0.17 & 6.5 $\pm$0.5 & -1.61$\pm$0.27 & 2.13$\pm$0.13 & 0.24$\pm$0.02 \\
BL Lacs (MA)                         & 163 & -             & -2.74$\pm$0.30 & 6.77$\pm$1.30             & -1.72$\pm$0.14 & 2.18$\pm$0.02 & 0.23$\pm$0.01 \\
\\
FSRQs (this work)                    & 161 & -0.11$\pm$0.06 & -2.22$\pm$0.09 & 5.1$\pm$2.0 & -1.62$\pm$0.46 & 2.52$\pm$0.08 & 0.17$\pm$0.02 \\
FSRQs (MA)                           & 161 & -             & -2.41$\pm$0.16 & 6.12$\pm$1.30            & -0.70$\pm$0.30 & 2.48$\pm$0.02 & 0.18$\pm$0.01 \\
\enddata
\tablenotetext{a}{The correlation between photon index $\Gamma$ and Log flux $F_{100}$.  See Equation \ref{fdef} and \S \ref{corrs}.  A higher value of $\beta$ (i.e. more positive correlation between flux and photon index absolute value) moves the mean of the photon index distribution down to lower photon index absolute value (lower $\mu$) and makes the faint end source counts slope less steep (less negative $m_{below}$), while a lower value of $\beta$ has the opposite effect.  All values reported for this work include the full range of results and their uncertainties when considering the 1$\sigma$ range of $\beta$. }
\tablenotetext{b}{The power law of the intrinsic flux distribution $\psi\!(F_{100})$ at fluxes above the break in the distribution.  See Equation \ref{bpowlaw}.    }
\tablenotetext{c}{The flux at which the power law break in $\psi\!(F_{100})$ occurs, in units of $10^{-8}$  photons cm$^{-2}$ sec$^{-1}$.  We present the value even though the precise location of the break is not important for the analysis in this work.  The value of $F_{break}$ can be obtained by equations \ref{normeqn} and \ref{normeqn2}. }
\tablenotetext{d}{The power law of the intrinsic flux distribution $\psi\!(F_{100})$ at fluxes below the break.  See Equation \ref{bpowlaw}. }
\tablenotetext{e}{The mean of the Gaussian fit to the intrinsic photon index distribution $h\!(\Gamma)$. }
\tablenotetext{f}{The 1$\sigma$ width of the Gaussian fit to the intrinsic photon index distribution $h\!(\Gamma)$. }
\tablenotetext{g}{Including all FQRQs, BL Lacs, and 28 of unidentified type.    }

\end{deluxetable}

\clearpage

\appendix

Here we test the methods used in this paper with a simulated Monte Carlo data set provided by the {\it Fermi}-LAT collaboration.  The data set is a realization of a set of simulations discussed in MA.  For each Monte Carlo realization 20000 sources were placed isotropically on the sky according to assumed distributions of flux (broken power law) and photon index (Gaussian).  Instrumental and observational effects on detection were applied to this data, resulting in a catalog of 486 sources with detection $TS$ of at least 50,  where $TS$ is the test statistic as discussed in \S \ref{datasec}.  

\begin{figure}
\includegraphics[width=3.5in]{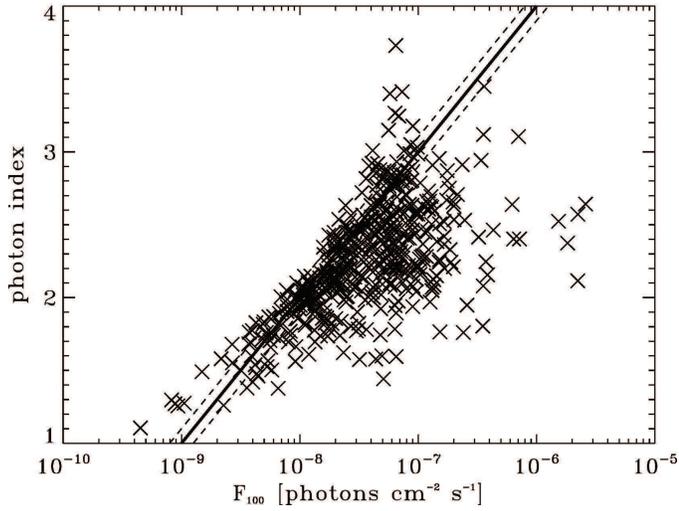}
\caption{ Flux and photon index for the 486 sources in the simulated {\it Fermi}-LAT data set, along with the curve (solid line) used to specify the observation truncation in the $F_{100}$,$\Gamma$ plane.  As with the real blazar data, moving the cutoff to the left dashed line has a large effect on the results, but moving it to the right dashed line has a negligible effect.}
\label{simlumsandsis}
\end{figure}

Figure \ref{simlumsandsis} shows the fluxes and photon indexes for the simulated data set, along with a curve approximating the observation truncation in the $F_{100}-\Gamma$ plane, which we take as the limiting flux for any object at a given photon index, and the limiting photon index for any object at a given flux, for inclusion in the relevant associated sets.  Since there is some uncertainty in the detection threshold values of fluxes and indexes we carry our analysis first assuming the individual limiting fluxes for each source to be  $F_{lim}=F_{100} / \sqrt{TS/50}$, and then by using a simple curve to define the truncation boundary as shown in Figure \ref{simlumsandsis}. This is a more conservative assumption and few sources are excluded but it insures the completeness of the sample. We experiment with moving the curve to the right and down (eliminating more sources in the edges of the sample) until we do not notice any change in the result. As described below this reproduces the input data accurately. We carry out the same procedure for the real data.

\begin{figure}
\includegraphics[width=3.5in]{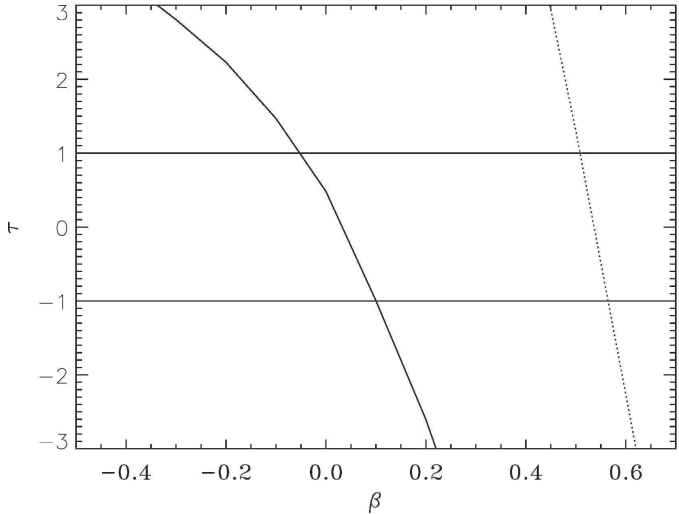}
\caption{ Correlation factor $\beta$ versus test statistic $\tau$ for a photon index and flux correlation of the form given in Equation \ref{fdef}, for the 486 sources in the simulated {\it Fermi}-LAT data set.  The solid curve show the results for the method employed here with the cutoff curve shown in Figure \ref{simlumsandsis} while the dotted curve shows the results for the raw data. }
\label{betazz}
\end{figure}

While the raw data obtained from these simulations show a strong correlation between flux and index (ignoring the truncation we obtain the  correlation parameter $\beta_{\rm raw}=0.53 \pm 0.03$ defined in equation \ref{fdef}), our method shows that once the effects of truncation are accounted for the correlation disappears and we get a correlation parameter $\beta$ consistent with zero, in agreement with that of the input data. Figure \ref{betazz} shows the values of $\beta$ vs $\tau$ for both the raw data and with the truncation accounted for.

\begin{figure}
\includegraphics[width=3.5in]{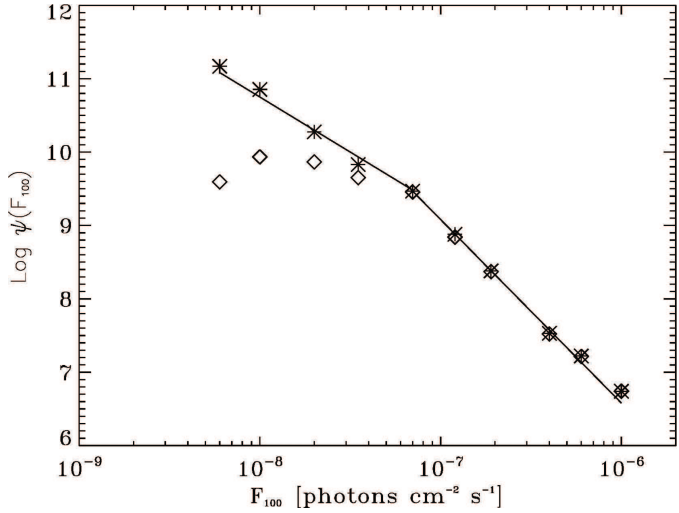}
\caption{ Observed (diamonds) and reconstructed $\beta=0$ intrinsic (stars) differential distribution of flux $\psi\!(F_{100})$ for the 486 sources in the simulated {\it Fermi}-LAT data set.  The data have intrinsic power law slope distributions of -2.49 and -1.59 above and below the break, respectively, which are plotted.  The normalization of $\psi\!(F_{100})$ here is arbitrary.  }
\label{simfluxdistfig}
\end{figure}

\begin{figure}
\includegraphics[width=3.5in]{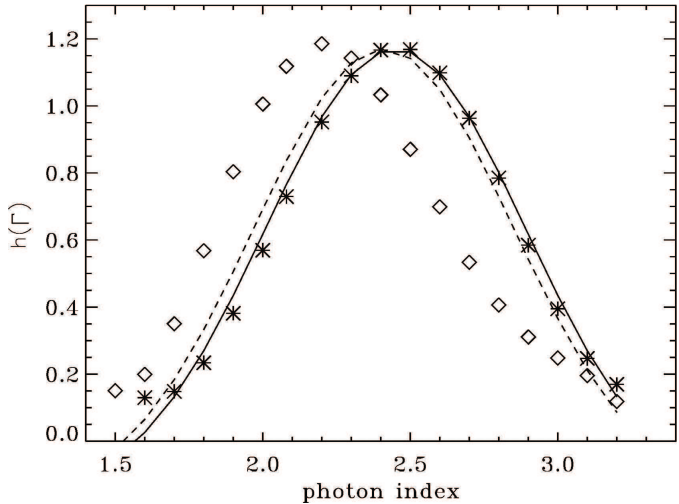}
\caption{ Observed (diamonds) and reconstructed $\beta=0$ intrinsic (stars) differential distribution of photon index $h\!(\Gamma)$ for the 486 sources in the simulated {\it Fermi}-LAT data set.  The data have an intrinsic Gaussian distribution with a mean of 2.37 and 1$\sigma$ width of 0.28, which is shown by the dashed curve.  The solid curve is the best fit Gaussian to the stars, which differs only slightly from the dashed curve.  The normalization of $h\!(\Gamma)$ here is arbitrary. }
\label{simadistfig}
\end{figure}

Figure \ref{simfluxdistfig} shows the observed and reconstructed intrinsic differential distribution of flux $\psi\!(F_{100})$ for the simulated data set, along with the known intrinsic distribution.  Figure \ref{simadistfig} shows the observed and reconstructed intrinsic differential distribution of photon index $h\!(\Gamma)$ for the simulated data set, along with the known intrinsic distribution.  Table 2 shows the full range of values for the reconstructed intrinsic distributions for the simulated data, including the effects of the entire 1$\sigma$ range of the correlation parameter $\beta$, along with the known input distributions.  It is seen that these methods successfully reproduce the input intrinsic distributions from a highly truncated data set.  

\begin{figure}
\includegraphics[width=3.5in]{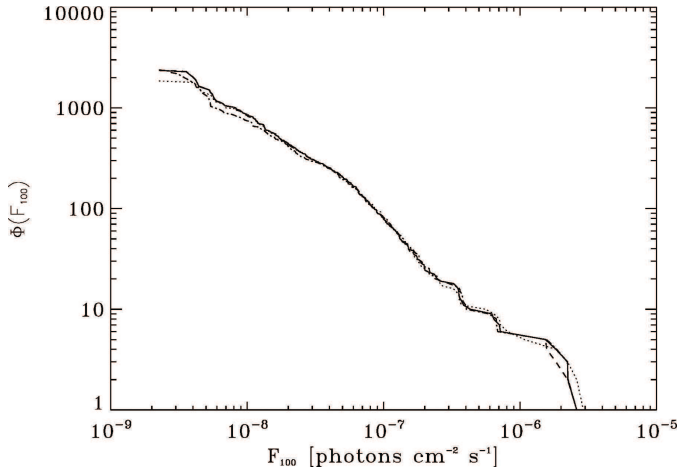}
\caption{Reconstructed $\beta=0$ intrinsic cumulative distribution of flux $\Phi\!(F_{100})$ for the 486 sources in the simulated {\it Fermi}-LAT data set. The normalization of $\Phi\!(F_{100})$ here is arbitrary.  The solid curve shows the fluxes as simulated, while the dashed (n=10) and dotted (n=100) curves show the results if a number n of those fluxes are altered in such a way that half of the altered fluxes are increased by 30\% and half are decreased by 30\%, values representing the largest typical reported uncertainties for the flux measurements used in this analysis.  The dash-dot curve shows the result for a more realistic case where the fluxes of all of the objects are altered in the manner described in the text.  As evident in all cases the added uncertainty introduced in the cumulative and fitted differential distribution is small compared to the uncertainty resulting from considering the extremal values of the correlation parameter $\beta$ or other uncertainties considered in this analysis.}
\label{simcumfluxdistfig}
\end{figure}

\begin{figure}
\includegraphics[width=3.5in]{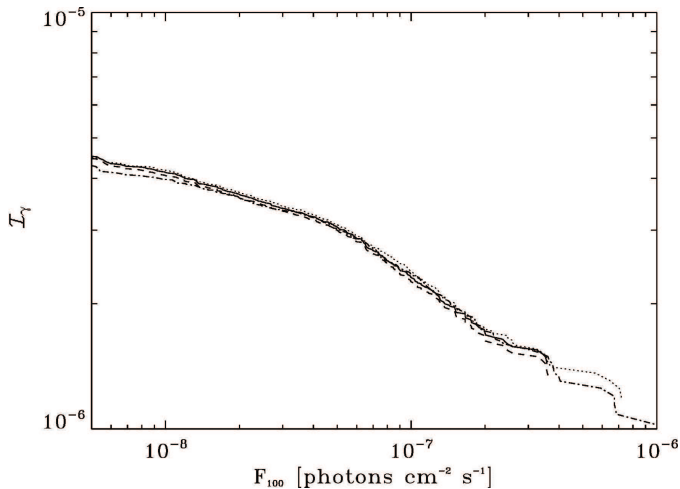}
\caption{Same as Figure \ref{simcumfluxdistfig} except the reconstructed $\beta=0$ estimate of the cumulative total number of photons between 0.1 and 100 GeV above a given $F_{100}$ from blazars, $\mathcal{I}_{\rm \gamma - blazars}(>F_{100})$, from equation \ref{CGB2}.   The normalization of $\mathcal{I}_{\rm \gamma - blazars}(>F_{100})$ here for the simulated data set is arbitrary. }
\label{simbgndfig}
\end{figure}

Here we can also address the question of whether errors in the cumulative distribution propagate in a significantly correlated way to the differential distributions.  First we take random samples of size n from the simulated data set and add 0.3 times the flux of each object to itself in half of each of the samples and subtract 0.3 times the flux of each object from itself in the other half.  This factor of 0.3 reflects the largest typical reported uncertainties in the {\it Fermi}-LAT flux measurements that we use.  The effect of doing so on the cumulative flux distribution is shown in Figure \ref{simcumfluxdistfig} while the effect propagated through to the cumulative total number of photons (i.e. the contribution to the EGB), determined in the manner of equation \ref{CGB2}, is shown in Figure \ref{simbgndfig}.  In both figures the solid curve shows the cumulative flux distribution $\Phi\!(F_{100})$ for n=0, the base case of no changes.  The dashed curves show n=10 and the dotted curves are for n=100, the later representing almost one quarter of the objects in the sample.  In these cases the differences in the cumulative flux distribution and the cumulative number of photons from the base case are negligible.  Then, for a more realistic but perhaps extreme case, we alter the flux of \underline{all} of the objects with alterations distributed such that those objects with the lowest fluxes have their fluxes altered by the highest typical reported measurement errors of 0.3 times the flux, while those with higher fluxes have lower errors, with the alteration proportional to the ratio of the difference of the logarithm of the flux and that of the maximum flux in the sample, with positive alterations for half the objects and negative alterations for half.  The resulting cumulative flux distribution and cumulative number of photons is shown by the dash-dot curves in Figures \ref{simcumfluxdistfig} and \ref{simbgndfig}.  Even then the change to the cumulative flux distribution is small and the added uncertainty introduced into the fitted differential distribution and the cumulative total number of photons if this case is considered relative to the base case is small compared to the uncertainty resulting from considering the extremal values of the correlation parameter $\beta$ and other sources of uncertainty considered in this work (compare with Figures \ref{fluxphifig} and \ref{bgndcontfig}).

We have also examined the effect of increasing the data set size on the uncertainty range determined for the correlation parameter $\beta$ for the method employed in this work.  A second simulated data set provided by the {\it Fermi}-LAT collaboration consisting of a catalog of $\sim$6 times as many (3018) objects resulted in a 1$\sigma$ range for $\beta$ that was approximately half as large ($\beta$=-0.01$\pm$ -0.04) as with the 486 object set.  As the uncertainty range in the value of $\beta$ is the major driver in the total uncertainty of the fitted distribution parameters, we can expect a significant reduction in uncertainty levels for the distribution parameters of the real data with a future five year {\it Fermi}-LAT extragalactic catalog consisting of $\sim$1500 blazars as opposed to the 352 here.

\begin{deluxetable}{lccccccc}\label{apptizzable}
\tabletypesize{\scriptsize}
\tablecaption{Simulated blazar intrinsic distributions, as calculated in this work and as known from the inputs to the simulated dataset} 
\tablecolumns{8}
\startdata
   & $\beta$\tablenotemark{a}  & $m_{above}$\tablenotemark{b} & $F_{break}$\tablenotemark{c} & $m_{below}$\tablenotemark{d} & $\mu$\tablenotemark{e} & $\sigma$\tablenotemark{f} \\
\hline
\\
Determined distributions\tablenotemark{g} & 0.02$\pm$0.07 &-2.41$\pm$0.11 & 8.0$\pm$1.0 & -1.61$\pm$0.27 & 2.38$\pm$0.9 & 0.34$\pm$0.05 \\
Known input distributions\tablenotemark{g}        &  0                       & -2.49 & 6.6 & -1.59 & 2.37 & 0.28 \\
\enddata
\tablenotetext{a}{The correlation between photon index $\Gamma$ and Log flux $F_{100}$.  See Equation \ref{fdef} and \S \ref{corrs}.}
\tablenotetext{b}{The power law of the intrinsic flux distribution $\psi\!(F_{100})$ at fluxes above the break in the distribution.  See Euqation \ref{bpowlaw}.   }
\tablenotetext{c}{The flux at which the power law break in $\psi\!(F_{100})$ occurs, in units of $10^{-8}$  photons cm$^{-2}$ sec$^{-1}$. }
\tablenotetext{d}{The power law of the intrinsic flux distribution $\psi\!(F_{100})$ at fluxes below the break.  See Equation \ref{bpowlaw}. }
\tablenotetext{e}{The mean of the Gaussian fit to the intrinsic photon index distribution $h\!(\Gamma)$.  For the analysis here this includes the full range of results and their uncertainties when considering the 1$\sigma$ range of $\beta$. }
\tablenotetext{f}{The 1$\sigma$ width of the Gaussian fit to the intrinsic photon index distribution $h\!(\Gamma)$. }
\tablenotetext{g}{The simulated observed blazar data set is provided by the Fermi collaboration, and has 486 objects.  All values for the determined distributions reported here include the full range of results and their uncertainties when considering the 1$\sigma$ range of $\beta$. }

\end{deluxetable}

\end{document}